\documentclass[aps,pra,twocolumn,twoside,notitlepage,amsmath,amssymb,amsfonts]{revtex4-1}
\usepackage{textcomp,hyperref,enumerate, url, cancel}                    			
\usepackage[margin=0.8in]{geometry}
\usepackage{color} 
\usepackage{amsmath,amssymb,amsthm,bbm,bm,nicefrac}   				
\usepackage{graphicx,epsfig}                                                               		
\usepackage{multirow, comment,subfigure}                                                             		

\newcommand{\ket}[1]{\left|#1\right\rangle}

\newcommand{\be}{\begin{equation}} 							
\newcommand{\ee}{\end{equation}}
\newcommand{\ba}{\begin{align}}
\newcommand{\ea}{\end{align}}
\newcommand{\bematrix}{\left(\begin{matrix}}
\newcommand{\ematrix}{\end{matrix}\right)}

\theoremstyle{definition}

\theoremstyle{theorem}

\theoremstyle{lemma}

\theoremstyle{proposition}

\theoremstyle{corollary}

\theoremstyle{observation}

\theoremstyle{remark}

\usepackage{algorithm}
\usepackage{tabularx}
\makeatletter
\def\BState{\State\hskip-\ALG@thistlm}
\makeatother
\usepackage{algpseudocode}
\makeatletter
\algnewcommand{\LineComment}[1]{\Statex \hskip\ALG@thistlm \(\triangleright\) #1}
\makeatother

\makeatletter
\newcommand{\multiline}[1]{%
  \begin{tabularx}{\dimexpr\linewidth-\ALG@thistlm}[t]{@{}X@{}}
    #1
  \end{tabularx}
}
\makeatother

\usepackage{algpseudocode}

\def\one{{\mbox{$1 \hspace{-1.0mm}  {\bf l}$}}}

\def\Z{\ensuremath{\mathbbm{Z}}}

\def\cA{\mathcal A}											
\def\cB{\mathcal B}
\def\cC{\mathcal C}

\def\cE{\mathcal E}

\def\cH{\mathcal H}

\def\cP{\mathcal P}

\def\cR{\mathcal R}
\def\cS{\mathcal S}

\newcommand\defn[1]{\textsl{#1}} 								

\begin{document}
\title{Reinforcement learning for optimal error correction of toric codes}
\author{Laia Domingo Colomer, Michalis Skotiniotis, and Ramon Mu\~noz-Tapia}
\address{F\'isica Te\`orica: Informaci\'o i Fen\`omens Qu\`antics, Departament de F\'isica, Universitat Aut\`onoma de Barcelona, 08193 Bellatera (Barcelona) Spain}
\date{\today}
\begin{abstract}
We apply deep reinforcement learning techniques to design high threshold decoders for the toric code under uncorrelated noise.  
By rewarding the agent only if the decoding procedure preserves the logical states of the toric code, and using deep convolutional 
networks for the training phase of the agent, we observe near-optimal performance for uncorrelated noise around the theoretically 
optimal threshold of $11\%$.  We observe that, by and large, the agent implements a policy similar to that of minimum weight perfect 
matchings even though no bias towards any policy is given \defn{a priori}.    
\end{abstract}
\maketitle

\section{Introduction}
Computation using quantum mechanical systems holds much promise as the ability of quantum systems to exist in 
exotic states, such as quantum entanglement and superposition, is known to yield significant advantages in 
computation~\cite{Galindo2002, Georgescu2014}, communication~\cite{Gisin2007}, and sensing~\cite{Giovannetti2011,Degen2017}. 
Due to the fragility of such exotic states to environmental decoherence, the ability to actively protect sensitive quantum information
against noise by quantum error correction~\cite{Gottesman2010,Terhal2015} is indispensable on the road to a fully fault-tolerant
quantum computing infrastructure~\cite{Campbell2017}.  Quantum error correcting codes (QECC) need to be efficiently 
implementable both in terms of the physical operations as well as the time needed to recover corrupted quantum data. A promising 
platform are topological error correcting codes (TECC)~\cite{Kitaev:98, Bombin:13} for which the recovery operations consist of
quasi-local error-syndrome measurements and local Pauli correction operations.  A drawback of QECC, and TECC
is that the error configuration space grows prohibitively large with the number of errors, and error syndromes exhibit high degeneracy 
making the design of optimal, fast decoders a highly non-trivial task. 

In recent years several approaches based on cellular-automata~\cite{Herold2015,Herold2017,Lang2018,Kubica:19}, 
renormalization group~\cite{Duclos2010a,Duclos2010b}, restricted Boltzman machines~\cite{Torlai2017}, and 
machine learning~\cite{Varsamopoulos:17,Krastanov2017,Baireuther2018,Fosel:18,Chamberland:18,Sweke:18, 
Liu:19,Varsamopoulos:19,Maskara:19,Andreasson:19} have produced a plethora of high performance QECC and TECC decoders.
Particularly for TECC decoders designed using these techniques have shown to achieve similar performance to the best known 
decoder based on the minimum weight perfect matchings (MWPM) algorithm~\cite{Edmonds1965}.
Specifically,~\cite{Torlai2017} used a stochastic neural network in a supervised learning paradigm to design decoders for the toric code under 
phase flip noise, whereas~\cite{Krastanov2017} used deep neural networks to design decoders that outperform MWPM for surface codes in the 
case of correlated noise. Baireuther \emph{et al.}~\cite{Baireuther2018} used a recurrent neural network, trained only on experimental data, to 
decode a surface code under correlated noise with high accuracy, whereas~\cite{Varsamopoulos:17} 
used feedforward neural networks to construct fast, but not necessarily optimal, decoders for surface codes. 
Shortly after the same authors showed that a combination of renormalization techniques with neural networks is capable of providing 
fast, high threshold decoders~\cite{Varsamopoulos:19}.   Feedforward neural networks were also used to construct high threshold decoders for
the toric and color codes under several noise models, including spatially correlated noise~\cite{Maskara:19}.

In the above machine learning approaches the training is performed in a supervised manner. A more general framework of machine learning is 
reinforcement learning (RL) where the agent is unsupervised and learns by simply interacting with its immediate environment and receiving 
feedback in terms of rewards.  Sweke \emph{et al.} combined RL with deep Q learning
(DQL) to design fast decoders for the surface code under correlated noise and noisy syndrome measurements, 
whereas in~\cite{Andreasson:19} RL and DQL were used to design 
high threshold decoders for the toric code under phase flip noise.  It is worth noting that machine learning techniques have also been used for
optimally designing the requisite QEC~\cite{Fosel:18, Nautrup:18,Valenti:19}.   
 
Here we construct model-free optimal decoders for the toric code in the presence of 
uncorrelated noise.  Just as in~\cite{Andreasson:19}, we use deep convolutional networks and episodic memory to train 
an agent in a RL paradigm but employ a fundamentally different system of rewards.  In~\cite{Andreasson:19}
agents were rewarded based on the number of actions taken, not whether the resulting actions yielded a successful decoding procedure. 
Indeed, the authors state that whilst the latter reward system is more natural, they found it difficult to train their agents using it.  
We show that agents can be trained within a few hours using rewards based solely on the correct decoding of the quantum information
and show that our decoders reach near optimal performance just shy of the optimal threshold. Our decoders exhibit good 
performance even when trained with error rates slightly above percolation.  We compare our agents performance to those of~\cite{Andreasson:19} 
by comparing their respective episode lengths and find them to be nearly identical thus further supporting the intuition that the optimal decoders
are those that take the minimum number of actions in order to correct the syndrome.
   
The article is structured as follows.  In Sec.~\ref{sec:TQEC} we briefly review the principles of QECC and the stabiliser formalism 
and introduce TECC and in particular the toric code.  In Sec.~\ref{sec:RL} we introduce the interaction-based learning scenario 
between an agent and an environment and give a basic review  of reinforcement learning and its implementation.  In Sec.~\ref{sec:decoders} we formulate the decoding of the toric code in the presence of uncorrelated noise as a reinforcement 
learning problem and present the results of applying such decoders of up to $9\times 9$ lattices.  We summarize and conclude in 
Sec.~\ref{sec:conclusion}.
\begin{figure*}[ht!]
    \centering
    \subfigure[]{\label{fig:torus}\includegraphics[scale=0.475]{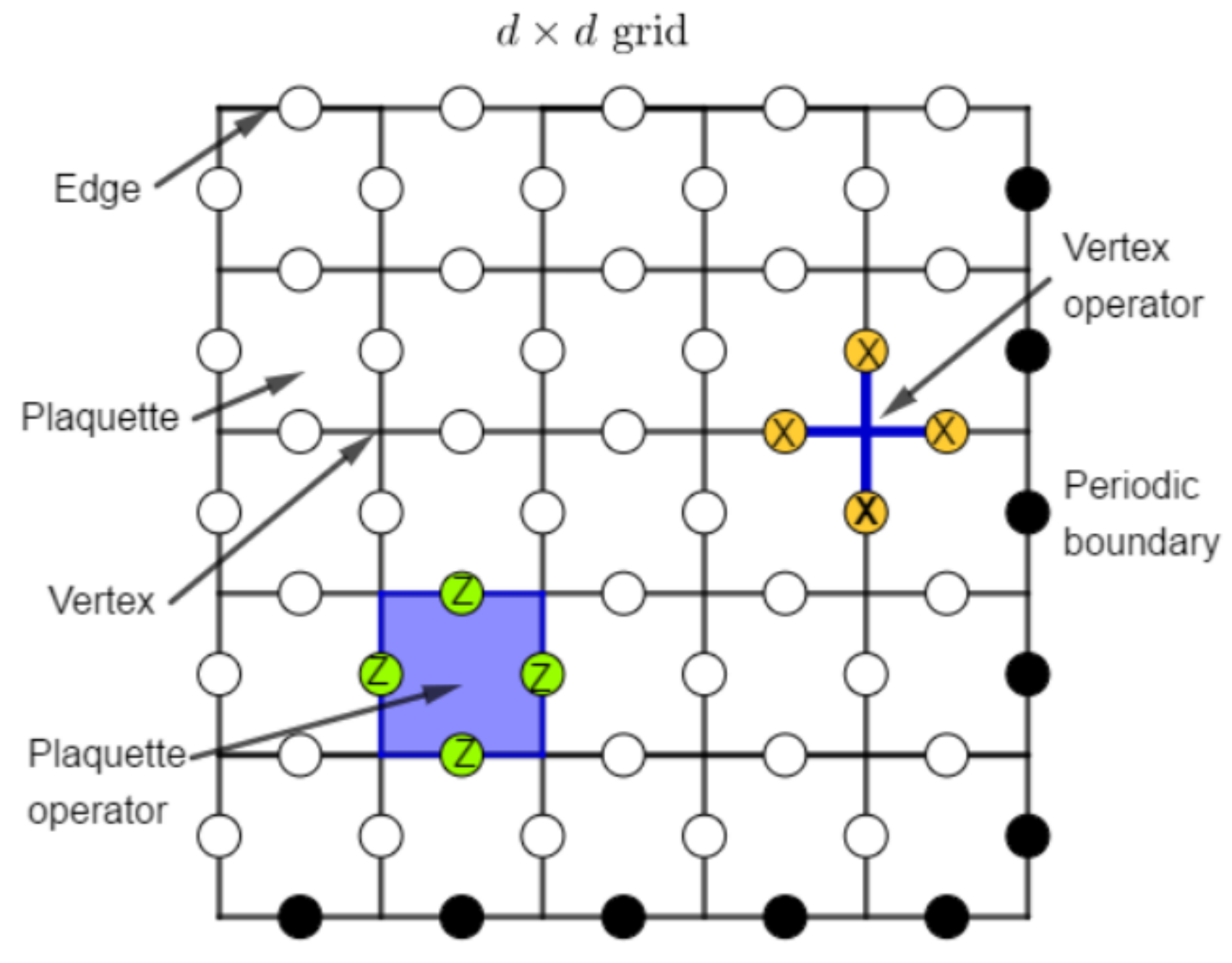}}\;
    \subfigure[]{\label{fig:trivial_loops}\includegraphics[scale=0.5]{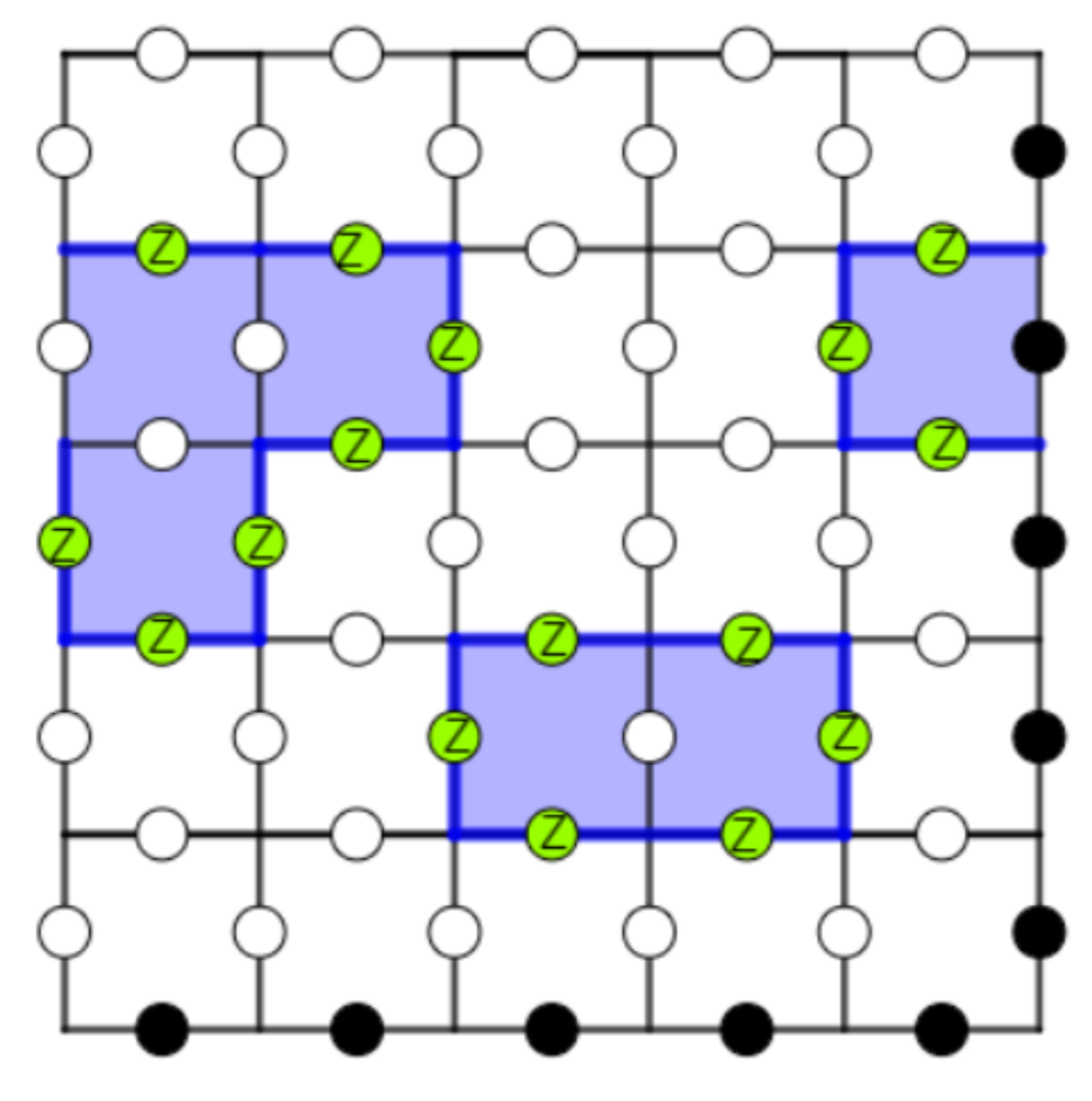}}
 \caption{The basic ingredients of the toric code: (a) A $5\times 5$ lattice with the topology of a torus.  The right and left boundaries of 
the lattice satisfy one periodic boundary condition, whereas the top and bottom edges the other periodic boundary condition.  
Physical qubits---represented as circles---are placed on the edges of the lattice.  The stabilizers of the toric code are associated with 
\emph{plaquette} operators (shaded blue square), and \emph{vertex} operators (dark blue star) (b)Examples of multiplication of 
plaquette operators. The resultant operator consists of $Z$ operators acting on the boundary of the combined plaquettes, and 
correspond to a stabilizer operator.}   
\end{figure*}
 
\section{Quantum error-correction and the toric Code}
\label{sec:TQEC}

Quantum information deals with the storage, transmission and manipulation of information represented in the states of quantum 
mechanical systems. Unfortunately, quantum systems are notoriously sensitive to the effects of noise which implies that their 
information depletes fairly quickly.  A way to counteract the deleterious effects of noise is to make use of quantum error 
correcting codes (QECC) (see~\cite{Lidar2013} and references therein).  Much like classical error correction the idea behind 
QECC is to use a number, $n$, of \emph{physical} quantum systems, each with an associated state space 
$\cH$, and identify a suitable subspace, $\cC\subset\cH^{\otimes n}$,  onto which quantum information can be protected by
decoherence.  A crucial ingredient in QECC are the encoding and decoding operations to and from the \defn{code space} $\cC$. 
Ideally we seek to design codes with large error tolerance, high storage capacity, and efficient encoding, decoding and recovery 
operations. Hereafter, all physical systems we consider are two-dimensional quantum systems (qubits).

The dimension, $d$, of the code space, $\cC$, defines the number of distinct \emph{logical} states, or codewords, as well as the 
number of \defn{logical} qubits, $k=\log_2 d$.  The \emph{distance}, $\delta$, of a code is the number of errors it can correct.  
By way of example, the three-qubit repetition code utilizes the code space 
$\cC:=\mathrm{span}\{\ket{0}_L\equiv \ket{000}, \; \ket{1}_L \equiv \ket{111}\}$, of three physical qubits to store one 
logical qubit and protect it against  a single qubit  $X$ error. Here we denote by  
\begin{equation}
X = \begin{pmatrix} 
0 & 1 \\
1 & 0 
\end{pmatrix},
\quad
Y = \begin{pmatrix} 
0 & -i \\
i & 0 
\end{pmatrix},
\quad 
Z = \begin{pmatrix} 
1 & 0 \\
0 & -1 
\end{pmatrix},
\label{eq:Paulis}
\end{equation}
the usual Pauli matrices.

As the number of physical qubits and error thresholds for QECC grows working directly with logical states and their superpositions
becomes inefficient.  Thankfully, an efficient description of QECCs exists in terms of the stabilizer 
formalism~\cite{Gottesman:97,Poulin:05}.  A subspace $\cC\subset\cH^{\otimes M}$ is said to be \defn{stabilized} by an operator
$P\in\cB(\cH^{\otimes M})$ if for any $\ket{\psi}\in\cC, \, P\ket{\psi}=\ket{\psi}$.  If this is the case then $P$ is called a \defn{stabilizer}
of the code and $\cC$ is uniquely specified as the eigenspace of the complete set of commuting stabilizers 
$\cP:=\{P_i, \, i\in(1,\ldots, N)\, |[P_i, P_j]=0, \forall\, i\neq j\}$ with eigenvalue $+1$.  Note that $\cP$ forms a finite Abelian group 
under matrix multiplication and consequently can be generated by $m=\log_2 N$ suitably chosen generators.  The number of generators $m$, 
logical qubits $k$,  and physical qubits $n$ are related by $m = n -k$.

In order to encode logical quantum information we need to construct the logical Pauli operators, $X_L, \, Z_L$ in such a way that they 
commute with the stabiliser group $\cP$.  For the three-qubit code, we have   
\be
\cP=\{Z\otimes Z \otimes \one, \quad Z \otimes \one \otimes Z, \quad \one \otimes Z \otimes Z, \quad \one \otimes \one \otimes \one\},
\label{eq:3qubitcodestabilizers}
\ee
and one can easily check that the following operators
\be
X_L = X \otimes X \otimes X \qquad Z_L = Z \otimes \one \otimes \one,
\label{eq:3qubit logical}
\ee
commute with those of Eq.~\eqref{eq:3qubitcodestabilizers}, and satisfy the Pauli commutation relations. 

Decoding, on the other hand, is a two-stage process involving first a recovery operation before extracting the relevant quantum 
information. The recovery operation consists of measuring all $2^m$ stabilizers and, based on the measurement 
outcomes---the \defn{error syndrome}---apply Pauli correction operations on the $n$ physical qubits.  If the syndrome contains all 
$+1$ then no recovery operation is required, whereas $-1$ values in the syndrome indicate the presence of errors.  
For the three-qubit code, $m=3-1=2$. The error syndromes of  $Z \otimes Z \otimes \one$ and $\one \otimes Z \otimes Z$ uniquely 
identify the physical qubit on which a Pauli $X$ error occurred.  Note, however, that in general the relationship between physical 
errors and the syndrome read-out  is not unique. There may be many error configurations which lead to the same syndrome, a 
phenomenon that occurs frequently in topological QECC which we now review.

\subsection*{Topological QECC: the toric code}
\label{sec:topologicalQEC}

Constructing QECC with high capacity for quantum information and large distance poses a serious challenge 
as both stabilizers and logical operations are generally \textit{global} operators acting on all $n$ physical qubits.  
An alternative, and more resource intensive, way of constructing QECCs is to exploit the topological properties of multi-qubit 
systems arranged on a lattice~\cite{Kitaev:98, Bombin:13}.  Such topological error correcting codes attain superb protection 
from decoherence, while requiring only local gates for error-correction and have been experimentally constructed in a variety of 
architectures~\cite{Yao:12, Nigg:14, Hill:15, Corcoles:15, Kelly:15, Riste:15, Takita:16}. In the remainder of this work, we shall focus 
on one of the simplest TECC, the toric code~\cite{Kitaev:98}. 

The toric code defined over an $d \times d$ square lattice, consists of $n=2 d^2$ physical qubits placed on every edge of the lattice. 
The topology of the torus arises from two distinct boundary conditions, one for the left and right edge of the lattice and one for the top 
and bottom edges. The stabilizer group of the toric code is generated by two distinct stabilizers associated with the 
plaquettes and vertices of the lattice (see Fig.~\ref{fig:torus}).  Specifically, to every plaquette, $p$, and vertex $v$, the operators 
\be
P_p=\hspace{-5mm}\bigotimes_{j\in\mathrm{boundary}(p)}\hspace{-5mm}Z_{j}, \qquad Q_v=\hspace{-3mm}
\bigotimes_{j\in\mathrm{vertex}(v)}\hspace{-3mm}X_{j}
\label{eq:plaquettesandstars}
\ee  
are stabilizers of the toric code, and there are a total of $2(d^2-1)$ independent plaquette and vertex operators forming the generators 
of its stabilizer group.  Observe that adjacent plaquette and vertex operators commute as they overlap on exactly two 
physical qubits. Products of plaquette (vertex) operators are also stabilizers of the torus and give rise to \emph{trivial loops} as 
illustrated in Fig. \ref{fig:trivial_loops}. 
\begin{figure}[h!]
\centering
    \includegraphics[scale=0.55]{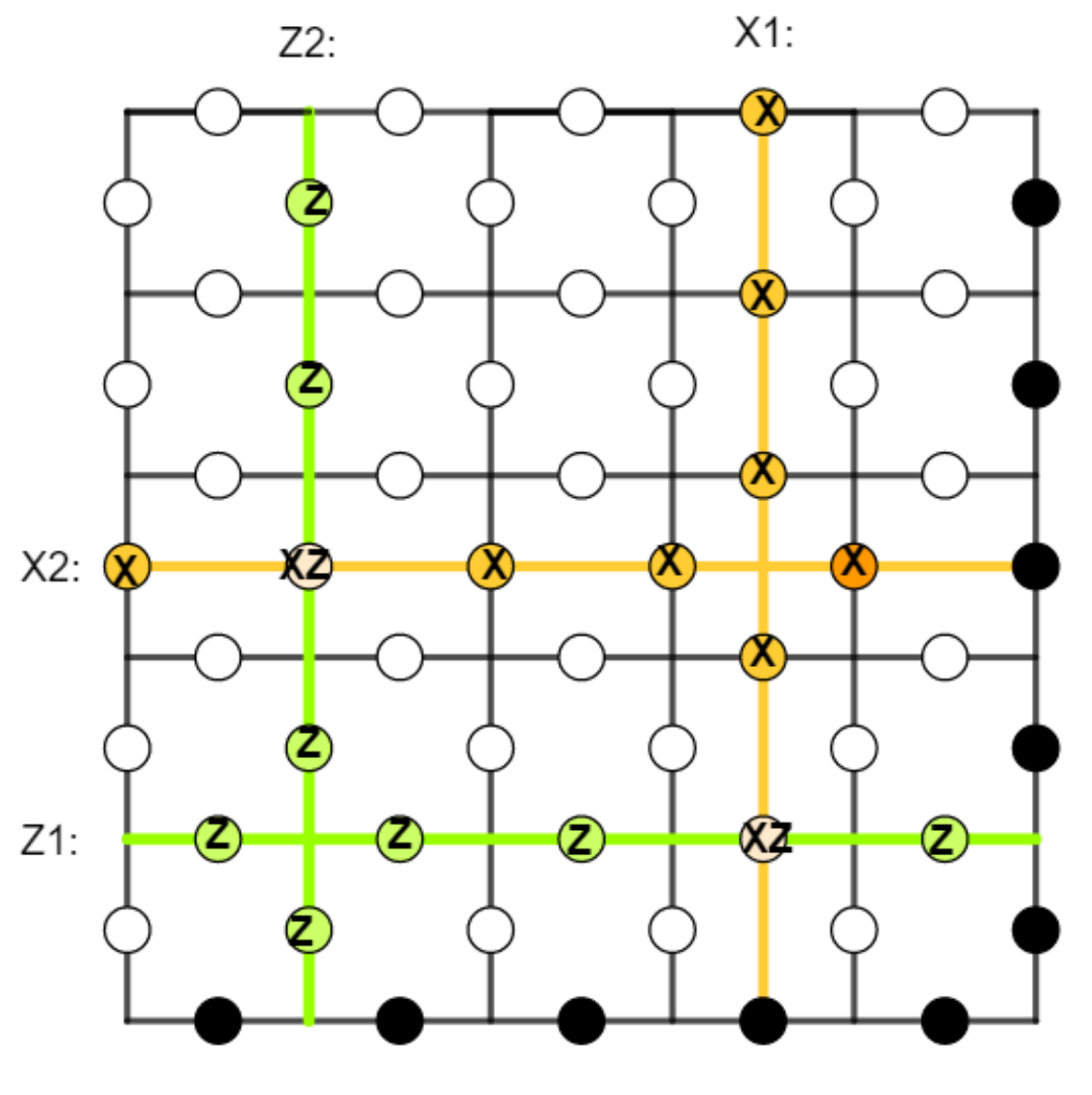}
     \caption {The logical Pauli operators $\{X_L^{(i)},\,Z_L^{(i)}\}_{i=1}^2$.}
     \label{fig:logicalXZ}
 \end{figure}

The dimension of the code space is $2d^2-2(d^2-1)=2$.  The logical operators, $\{X^{(i)}_L,Z^{(i)}_L\}^2_{i=1}$, for each of the logical 
qubits are shown in Fig.~\ref{fig:logicalXZ}.  They form \emph{non-trivial} closed loops around the torus with $Z^{(1)}_L$ forming 
horizontal loops, $Z_L^{(2)}$ vertical loops and the corresponding $X^{(i)}_L$ being closed loops orthogonal to those of $Z^{(i)}_L$.
Notice, however, that the construction of the logical operators is not unique: we can generate an equivalence class of logical operators, 
acting identically on the code space, by multiplying the above logical operators with elements of $\cP$ (i.e., the loops do not need to be straight).

\begin{figure}[h!]
\centering
    \includegraphics[scale=0.4]{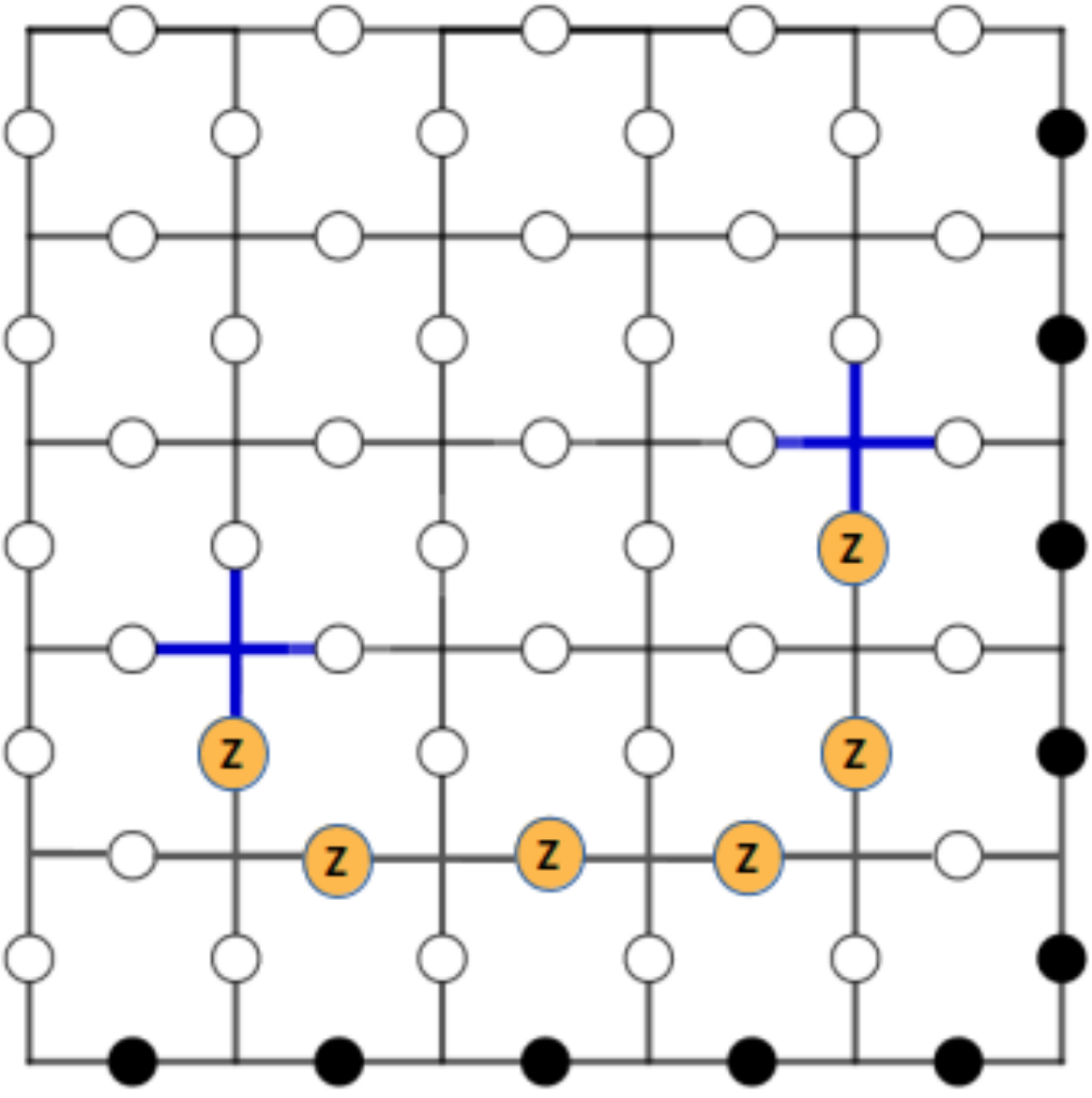}\, \includegraphics[scale=0.4]{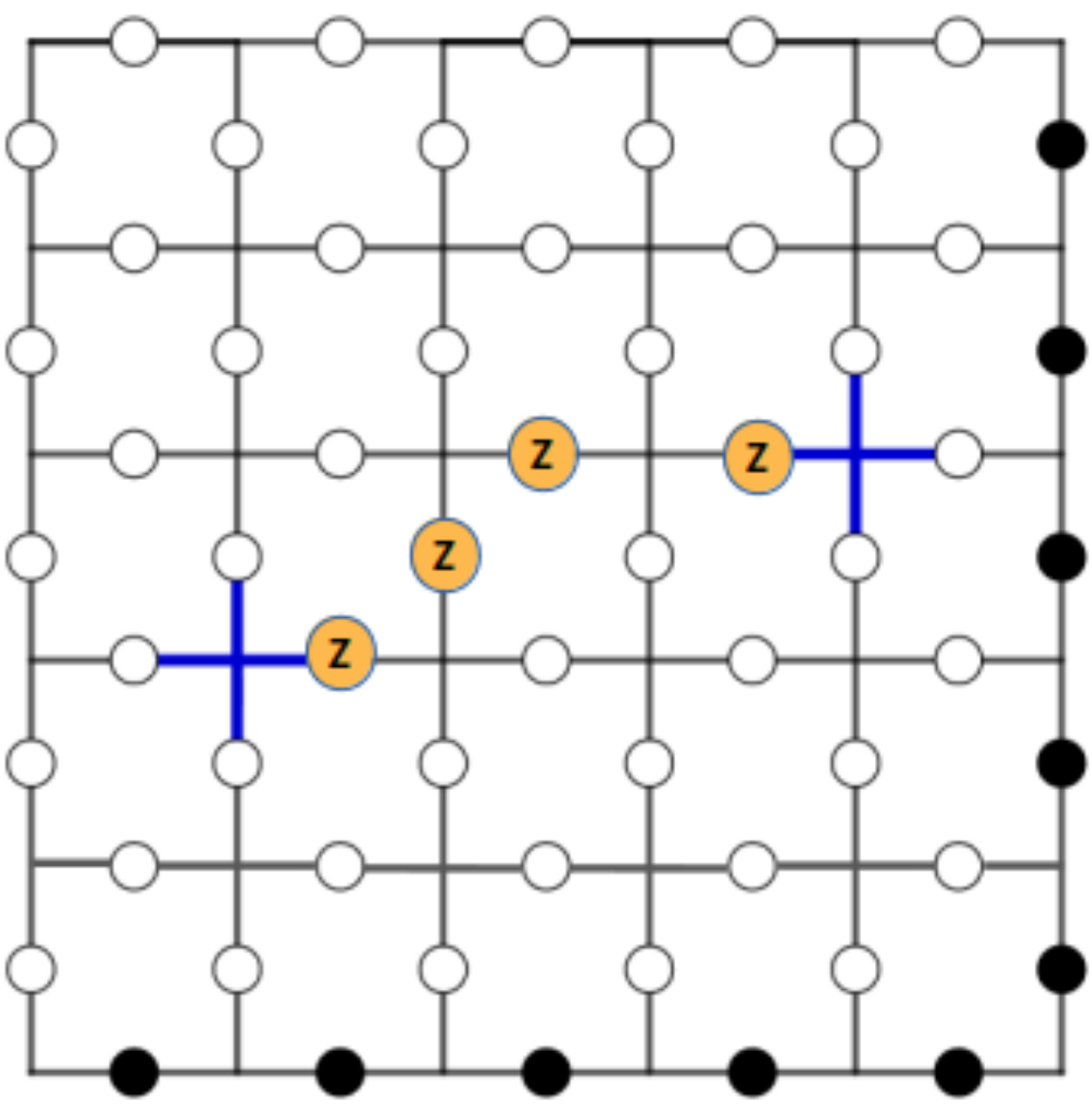}
     \caption {Example of different physical errors (orange circles) leading to the same vertex syndrome error (blue lines crossing a 
     vertex).}
      \label{fig:errors}
 \end{figure}

From the preceding discussion it follows that for a logical error to occur, an odd number of non-trivial loops around the torus must
occur.  Therefore, the distance of the toric code is $d$.  If a physical qubit suffers an error, the stabilizer generators 
adjacent to the position of the physical qubit will have error syndrome $-1$. However, the error syndrome distribution is never one to one with 
the physical errors of the lattice. There can be many different combinations of physical errors which lead to the same error syndrome, as shown in 
Fig.~\ref{fig:errors}. Therefore, smart strategies are needed to return the faulty state to the initial logical sate. The agent must perform physical 
operations which do not result in non-trivial loops around the torus, and thus logical errors, whilst returning the state to the codespace.

In addition, the design of optimal decoders also relies heavily on the types of errors occurring as well as their distribution. The simplest  
and most common noise models assume that each qubit experiences an independent and identically distributed (i.i.d.) noise process, 
with a probability $p$ of suffering an error.  Among the uncorrelated noise models, the most relevant ones are   
bit-flip ($X$ errors) and phase-flip ($Z$ errors) errors.  For \defn{correlated errors}, depolarizing noise ($X,\, Y,\, Z$ noise each with 
probability $\nicefrac{p}{3}$) is the most paradigmatic noise model.  In this work we shall only consider uncorrelated noise and without 
loss of generality we shall assume bit-flip noise (analysis for phase-flip errors is completely analogous).  
Correlated noise is more challenging and will 
be addressed in future work.  The optimal threshold of a decoder, on the other hand, is the 
maximum value of $p$ for which recovery of the information is possible.  For the 
toric code under i.i.d bit-flip errors this threshold is known to be $11\%$~\cite{Dennis:02}.

A widely used decoder for the toric code under bit-flip errors is based on the  minimum weight perfect matchings (MWPM)  
algorithm~\cite{Gabow:76,Cook:99}. MWPM adopts the policy of correcting for the most likely error given a particular error syndrome  
and has proven to be a very successful decoder with an estimated threshold of  $10.3\%$~\cite{Browne:14}.  Recently, machine learning 
techniques and applications of (deep) neural networks have been applied in search for optimal decoders, both for topological as 
well as standard QECC~\cite{Varsamopoulos:17,Krastanov2017,Baireuther2018,Fosel:18,Chamberland:18,Sweke:18, 
Liu:19,Varsamopoulos:19,Andreasson:19}.  
We now review the main techniques in reinforcement learning which we will use in search of efficient decoders for the toric code.

\section{Reinforcement learning}
\label{sec:RL}
Reinforcement learning (RL) is a framework within which one can precisely formulate the old dictum of ``learning through 
experience''~\cite{Sutton:1998}.  
Agents trained using RL have excelled at performing certain tasks, such as mastering the game of Go~\cite{Silver2016}, better than 
humans and RL based agents are used extensively in robotics~\cite{Orr2003}, artificial intelligence~\cite{Ramos2017}, and 
face-recognition~\cite{Brunelli:93}.  Here we introduce the agent-environment paradigm of RL and review its key features; state and 
state-action valued functions. We then briefly discuss deep Q-learning (DQL) which uses deep convolutional networks highlighting some key 
techniques used to guarantee convergence in the training process for the agent.

\begin{figure}[ht!]
\centering
\includegraphics[scale=0.3]{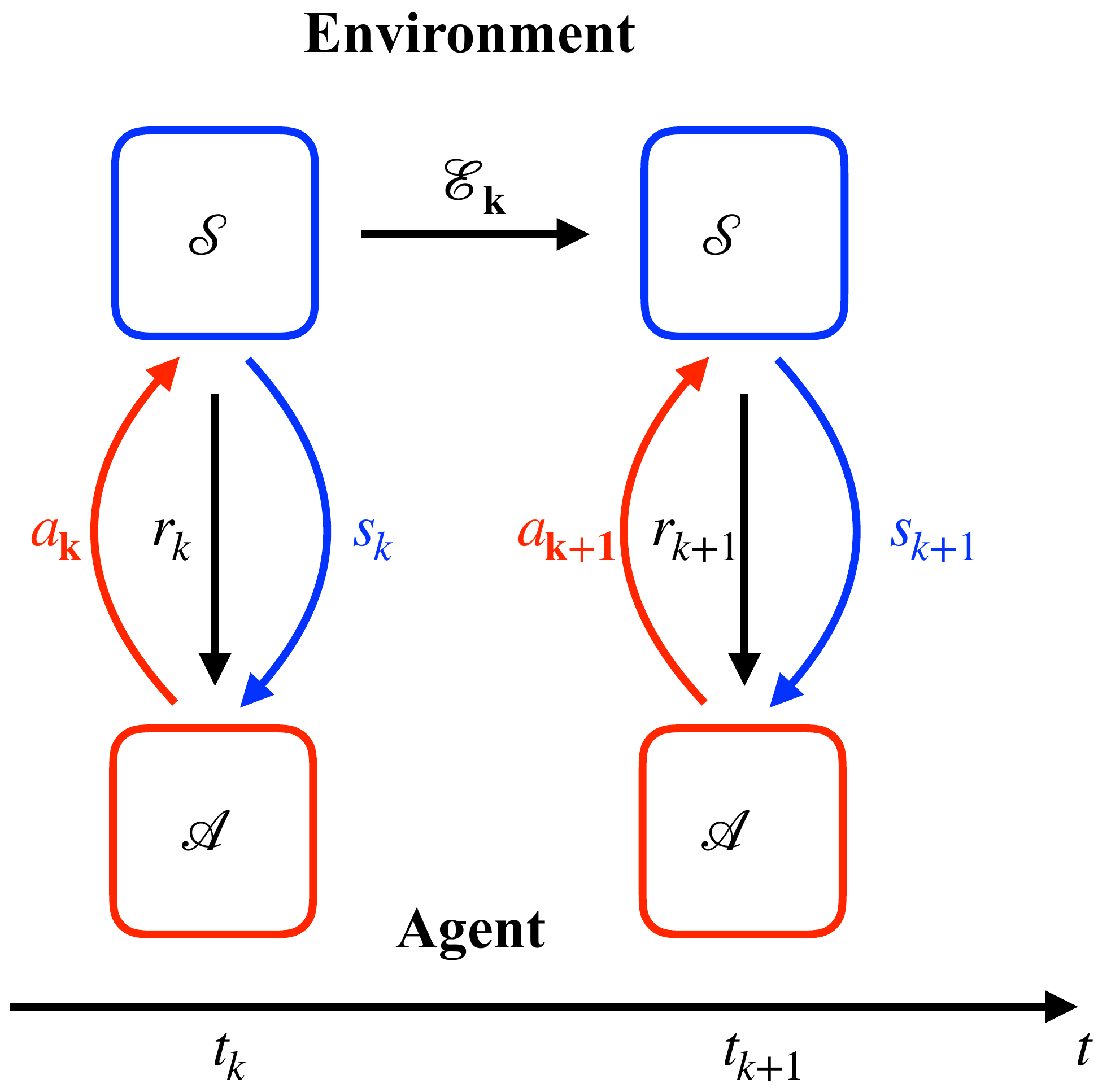}
\caption{A step in an episode in a general RL problem. The agent receives reward $r_k$ from her/his previous action while the current 
state of the environment is $s_k\in\cS$. The agent performs action $a_{\bf k}\in\cA$ which causes the environment to update its state to 
$s_{k+1}=\cE_{\bf k}(s_k)$, and reward the agent with $r_{k+1}\in\cR$. This interaction is repeated until a terminal state $s_T\in\cS$ is reached, 
and the episode finishes. The sequence  $\{(r_k,\, s_k, \, a_k)\}_{k=1}^T$ within the episode is called a \emph{trajectory}.}
\label{fig:Episode}
\end{figure}

Consider a scenario involving an agent, $A$, sequentially interacting with its immediate environment, $E$, in order to learn 
how to achieve a specific task (see Fig.~\ref{fig:Episode}).  Here, learning is to be understood as $A$'s ability to refine its future 
behaviour based on past experience in order  to maximise future reward. Regardless of the details of $A$, $E$, and their 
interaction any such learning scenario can be modelled using the following three ingredients~\cite{Sutton:1998}: the set of all possible 
states, $\cS$, of $E$, the set of all possible actions, $\cA$, of $A$, and the set of \defn{rewards}, $\cR$---an assessment of $A$'s 
performance towards the task.  If the interaction is known, then one talks of \emph{model-based} RL, otherwise the latter is \emph{model-free}.

The learning process of $A$ can be described in terms of \defn{episodes}~\footnote{Note that we can always consider finite sequences by introducing a terminal state for the environment indicating that the task is completed.}. $A$ and 
$E$ interact, as shown in Fig.~\ref{fig:Episode}, and the episode finishes when the agent reaches the terminal state. 
Any given step $k\in \Z_N$ in an episode consists of $A$ receiving the reward 
$r_k\in \cR$---from her/his previous action---and the current state of $E$, $s_k\in\cS$.  
$A$ then performs action $a_{\bf k}\in\cA$, after which $E$ sends reward $r_{k+1}\in\cR$, and 
its state changes to $s_{k+1}\equiv\mathcal{E}_{\bf k}(s_k)\in\cS$.  Here boldface subscripts for the actions and state changes to the state of the 
environment indicate that these may depend on the history of actions and past states of the environment.

The set of states, rewards, and actions of an episode are random variables, and the interaction between $A$ and $E$ 
is a stochastic map $\cE_{\bf k}:\cS\to\cS$, that may depend on all preceding states of $E$ and actions of $A$.  Specifically,  
the probability that $A$ receives reward $r_{k+1}$, and the state of $E$ at step $k+1$ is $s_{k+1}$,  given 
all preceding states $s_{\bm k}=s_1s_2\ldots s_k$ and actions $a_{\bm k}=a_1a_2\ldots a_k$ is
$p(s_{k+1}, r_{k+1}\lvert s_{\bm k},\,a_{\bm k})$. 
If the conditional probabilities depend solely on the last preceding state of $E$ and action of $A$, i.e., $p(s_{k+1}, r_{k+1}\lvert s_{\bm k},
\,a_{\bm k})=p(s_{k+1}, r_{k+1}\lvert s_k,\,a_k), \, \forall\, s_{k+1},\, s_k,\, r_{k+1},\,\mathrm{and}\, 
a_k$ then every finite sequence of steps  is formally equivalent to a finite Markov decision process 
(MDP)~\cite{Bellman:57}.  The average reward an agent $A$ expects to obtain after performing action $a_k$, given the state of the 
environment is $s_k$, is then 
\be
\mathbb{E}[r_{k+1}(s_k,a_k)]=\hspace{-2mm}\sum_{r_{k+1}\in\cR}\hspace{-2mm}r_{k+1}\hspace{-2mm}\sum_{s_{k+1}\in\cS}\
\hspace{-2mm}p(s_{k+1},r_{k+1}|s_k,a_k).
\label{eq:expectedreward}
\ee 
Note that whilst we have assumed that both state and action spaces are finite dimensional, Eq.~\eqref{eq:expectedreward} can be 
equally applied to infinite dimensional cases.

The learning of the agent is quantified by the \emph{expected discounted return}
\be
R_k:=\sum_{n=0}^T\gamma^n\, r_{k+n+1},
\label{eq:return}
\ee
where $0\leq \gamma\leq 1$ is the \emph{discount rate}~\footnote{The use of a discount rate is particularly useful for dealing 
with continuous interactions between agent and environment, i.e., for continuous sequences.}; $\gamma=0$ favours only immediate rewards, and for $0<\gamma<1$, larger values of $\gamma$ give more importance to long-term rewards than smaller values of $\gamma$. In the limit, when $\gamma=1$, all rewards are given the same value.
The decision making process of $A$ is known as a \emph{policy} and consists of a complete specification of the actions $A$ will
perform at every step of the sequence and for any possible state of $E$.   Given a state $s_k$ the probability that $A$ will 
perform action $a_k$ is denoted by $\pi(a_k|s_k)$.  Under a policy $\pi$  the value of a state $s_k$, $v_\pi(s_k)$, 
quantifies its average expected future returns., i.e., $v_\pi(s_k)=\mathbb{E}[R_k\vert s_k]_{\pi(\cA|s_k)}$.   Similarly, the value of an 
action $a_k$ given a state $s_k$---known as the \emph{q-value} $q_\pi(s_k,\,a_k)$--- quantifies the average 
expected future returns of that action, i.e.,  $q_\pi(s_k,\,a_k)=\mathbb{E}[R_k\vert s_k,\,a_k]_{\pi(\cA,\,a_k|\cS,s_k)}$. 
Such policy-value functions induce a partial order in the space of all 
possible policies of an agent: $\pi$ is at least as good as $\pi'$, if and only if $v_{\pi}(s_k)\geq v_{\pi'}(s_k),\,\forall\, s_k\in\cS$.  
A policy $\pi^*$ is optimal if no other policy can give a higher value than it, $v_{\pi^*}(s_k)=\max_{\pi} v_{\pi}(s_k)$~\cite{Sutton:1998}.

To determine the optimal policy one makes use of the recursive nature of both $v_\pi(s_k)$, and $q_\pi(s_k,a_k)$ to write 
\begin{align}\nonumber
v_{\pi}(s_k)&=\sum_{a_k\in\cA}\pi(a_k|s_k)\sum_{s_{k+1}}\sum_{r_{k+1}}p(s_{k+1},r_{k+1}|s_k,\,a_k)\\  \nonumber
&\times\left(r_{k+1} +\gamma v_{\pi}(s_{k+1})\right)\\
&=\sum_{a_k\in\cA}\pi(a_k|s_k) \mathbb{E}[r_{k+1}(s_k,a_k)] +\gamma v_{\pi}(s_{k+1})
\label{eq:state-value}
\end{align}
and 
\be
q_\pi(s_k,a_k)= \mathbb{E}[r_{k+1}(s_k,a_k)]+ \gamma q_\pi(s_{k+1},\,a_{k+1}).
\label{eq:action-value}
\ee
Eqs.~\eqref{eq:state-value}, and~\eqref{eq:action-value} are known as the \defn{Bellman} equations~\cite{Bellman:66} for state and 
state-action-value functions.  For the optimal policy $\pi^*$, $v_{\pi^*}(s_k)$ takes the specific form
\begin{align}\nonumber
v_{\pi^*}(s_k)&=\max_{a_k\in\cA}\sum_{s_{k+1}}\sum_{r_{k+1}}p(s_{k+1},r_{k+1}|s_k,\,a_k)\\
&\times\left(r_{k+1} +\gamma v_{\pi^*}(s_{k+1})\right)
\label{eq:Bellmanoptimality}
\end{align}
known as the Bellman optimality equation.  Note that an optimal policy is known to always exist, though it may not be 
unique~\cite{Puterman:14}.

\subsection*{Agent training: Deep Q-learning}
If the environment in a MDP is known then the optimal policy can be obtained by solving $\lvert\cS\rvert$ Bellman equations.
For model-free RL no such possibility exists and consequently state and state-action-value functions need to be estimated from experience.
A typical algorithm used in this case is called \emph{Q-learning}, with guaranteed convergence to the optimal $q$-value if every 
state-action pair is observed a sufficiently large number of times, i.e., if the agent is trained infinitely long~\cite{Sutton:1998}.  For large state 
spaces this is prohibitively expensive.  Consequently we have to resort to finite training sets which in turn means that the agent will 
often encounter situations previously unseen. 

Deep Q-learning (DQL) uses deep convolutional networks~\cite{Kalchbrenner2014}, specialised for processing high-dimensional data, in order 
to extract global features and patterns.  Upon encountering a previously unseen state, DQL  uses such global features to compare
with similar situations in past experience~\cite{Mnih:15}.  DQL  parametrises the $q$-function in terms of a neural network, so that 
given an input state and action, the neural network  produces the $q$-value $q(s,a)$ as an output. During training, the network 
parameters are adjusted, via stochastic gradient descent, such as to reduce the error between  the optimal and approximated 
target $q$-values. 

We use DQL to train our agent to successfully decode uncorrelated bit or phase flip noise on the toric code.  Training halts 
either after a certain number of episodes have happened, or until the loss function of the convolutional neural network stops decreasing.  To 
ensure stability during training we also make use of additional training techniques, such as double deep Q-learning, duelling deep 
Q-learning, and prioritised experience replay~\cite{Mnih2013, Simonini:18}. 
   
\section{Deep RL decoders for uncorrelated noise}
\label{sec:decoders}

We now cast decoders for the toric code under uncorrelated noise as a RL problem and present the results of training model-free 
agents to accomplish the task.  As already mentioned we discuss only the case of bit-flip errors. 
The environment, $E$, consists of the state of the toric code; a matrix of $2d^2$ entries containing the position of errors applied to the 
physical qubits for any given episode.  This state is hidden from $A$ and it is used to generate the state space $\cS$, comprised of  
the error syndrome of all stabilisers $\cP$ of the code, in this case a set of $d \times d$ matrices representing the position of each 
stabilizer and its corresponding error syndrome. This situation can be regarded as an example of a partially observable Markov decision 
process~\cite{barry2014}.
 
The agents actions consist of single qubit bit flip operations.  In principle, we can allow the agent to act on any of the $2d^2$ qubits.  
However, it is inconvenient to let the agent perform bit-flip operations to qubits which do not have an adjacent violated syndrome. 
Such qubits are not candidates of having produced the syndrome errors, and thus should not be altered. Therefore, for a given state $s$, i.e., 
a given error syndrome, only a subset of all possible actions $a$ (single qubit bit-flips) should be allowed and these actions are different 
for different syndrome errors. As the neural network needs to have a fixed output dimension, we need to make adjustments in order to appropriately 
accommodate the restrictions on the agents possible actions. One possibility would be to calculate the 
q-values for all $2d^2$ qubits, and allow the agent to perform only the valid actions but this results in large training times for the agent. 
Whilst this strategy does work it is extremely inefficient, particularly for large lattice dimensions.

\begin{figure}[h!]
\centering
    \includegraphics[scale=0.35]{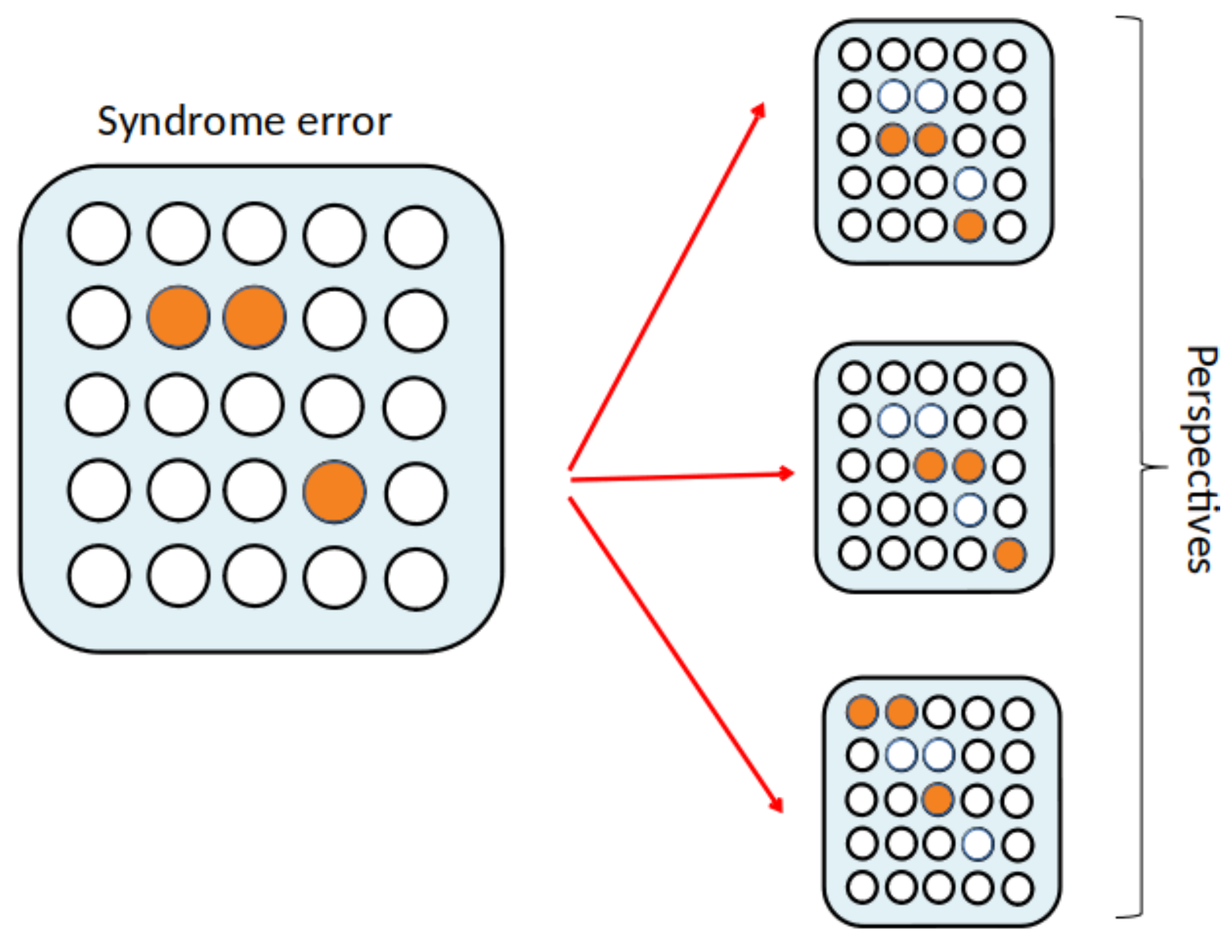}
     \caption {Example of the representation of errors given to the neural network. Given a syndrome with $N$ errors, we produce $N$ 
     perspectives 
     by locating a defect to the central plaquette, while keeping the relative position of the other defects fixed. The set of perspectives are 
     given to 
     the neural network as an input.}
      \label{fig:Perspectives}
 \end{figure}

In order to speed up the learning stage, a convenient representation of the errors must given to the neural network. 
By exploiting the 
boundary conditions of the torus Andreasson \emph{et al.} provide one such representation in terms of so-called 
\emph{perspectives} shown in Fig.~\ref{fig:Perspectives}~\cite{Andreasson:19}. The syndrome error of any stabilizer can be represented as an 
arbitrary plaquette at the center of the torus. For each syndrome we can generate a set of matrices---the perspectives $B_i$--each of them 
having a different defect at its center while keeping the relative position of other defects fixed.  The input to the neural network are the 
perspectives $B_i$, for which the neural network provides the $q$-value for each of the four possible actions; $X$ corrections on 
the qubits adjacent to the vertex in question. In other words, instead of inputing one single matrix containing the syndrome errors, $N$ matrices 
are given as an input, $N$ being the number of syndrome errors. In this way, the output of the neural network is reduced from $2d^2$ to $4N$, 
allowing for a more efficient and stable training phase.

The agent continues to perform actions until the \emph{terminal state} of the environment is reached: all syndrome measurements 
have outcome $+1$.  As we compute the actions $A$ performs on the hidden state of the code along the way we can evaluate the 
number of non-trivial horizontal or vertical loops around the torus.  If an even number of such loops is found, no logical errors have 
occurred and the agent is rewarded a nominal reward of $r=+1000$~\footnote{The value of the reward is chosen so as to 
speed-up the training process}.  Else, the agent's reward is $r=0$.  In~\cite{Andreasson:19}, a RL
decoder  for the same task was designed using similar techniques as ours.  There the agent was penalised with $r=-1$ for 
every iteration, so that the optimal strategy would be to correct the error syndrome with the minimum number of operations.  
We shall refer to the decoder of~\cite{Andreasson:19} as a \emph{minimum action decoder}, or MAD for short, and will compare 
this reward scheme with ours.
 
We train the agent using the DQL algorithm, until the parameters of the convolutional neural network stabilize. 
In order to give the agent more freedom to explore the large policy space during training we use an $\epsilon$-greedy strategy for 
$Q$-learning;  with probability $1-\epsilon$ the agent selects the action with the highest value of the current $q$-function, whereas with 
probability $\epsilon$ the agent performs an action at random.  Furthermore, we vary the value of $\epsilon$ during training, linearly 
decreasing it to the minimum value of $\epsilon=0.1$.  Moreover, we also train the agent with an initially 
low probability of bit flip errors so that $A$ learns to correct properly, linearly increasing the occurrence of errors near-to and beyond the $11\%$ 
threshold of the code. 

\subsection*{Neural Network architecture and training parameters}

In this section we provide the technical details used to design the neural network and training phase of the agent. The neural network 
consists of some convolutional layers, a set of fully connected layers and an output layer with four values.  Each output corresponds to applying 
an action to one of the qubits adjacent to the input plaquette.  The algorithm we implement is given in ~Table~\ref{DeepQ_alg}.  
The network is trained by adjusting the hyper parameters $\theta_i$ at iteration $i$ such that  the error between the optimal target values 
$r + \gamma \max_{a'} q^*(s',a')$ and the approximated target values $y_i = r + \gamma \max_{a'} q(s',a', \theta_i)$, calculated using the Q-network, is 
reduced.  Training consists in decreasing this loss function via stochastic gradient descend, until the Q-network produces precise values of the 
q-function. 

The fact that the approximate target values $y_i$ depend on the network parameters  produces instabilities in the training process, and could lead to 
divergence of the network weights. To that end we use a separate network for generating the targets $y_i = r + \gamma \max_{a'} \hat{q}(s',a', \theta_i)$ 
in the Q-learning update. More precisely, for every $K$ updates we clone the \textit{active} Q-network (the one used to select the best action in each 
state) to obtain a \textit{target} Q-network, which is used to generate the targets $y_j$ for the following $K$ updates.  Table~\ref{NN_architecture} shows 
the parameters used for the neural network architecture. Table~\ref{common_ hyper} shows the common hyper-parameters used for the agents, whilst 
table~\ref{single_ hyper} shows the hyper-parameters which were different.

\begin{algorithm}[H]
\caption{Deep Q-learning }\label{DeepQ_alg}
\begin{algorithmic}[1]
\Procedure{Initialization:}{} 
\State Initialize replay memory $D$
\State Initialize active Q-network $q$ with random weights $\theta$
\State Initialize target Q-network $\hat{q}$ with random weights $\theta^-$
\vspace{0.2in}
\For{t=1 $\cdots$ M}
\vspace{0.1in}
\State \multiline{With probability $\epsilon$ select a random action $a_t$}
\State \multiline{Otherwise select action $a_t = \text{argmax}_{a'} q(s',a,\theta)$}
\State \multiline{Execute action $a_t$, receive reward $r_t$ and next state $s_{t+1}$}
\State \multiline{Store experience $e_t = (s_t, a_t, r_t, s_{t+1})$ in memory $D$}
\State \multiline{Select a random sample $\{e_{i_1}, \cdots , e_{i_j}\}$ from D}
\vspace{0.05in}
\For{every experience in sample}
\vspace{0.1in}
\State \hspace{-0.2in}\multiline{Calculate target values $y_i = r_i + \gamma \max_{a'} \hat{q}(s_{i}, a_i, \theta^-)$, with $\hat{q}(s_{i}, a_i, \theta^-) =0$ if $s_i$ is a terminal state}
\vspace{0.1in}
\State \hspace{-0.2in}\multiline{Perform a gradient descent step on $(y_i - q(s_i, a_i,\theta))$ with respect to the network parameters $\theta$:}
\State \hspace{-0.2in}$\Delta \theta = \alpha(r_i + \gamma \max_{a'} \hat{q}(s_{i+1}, a', \theta^-) - q(s_{i}, a_{i}, \theta)) \nabla_\theta q(s_i, a_i, \theta) $
\vspace{0.1in}
\EndFor
\State Every $K$ iterations reset $\hat{q} = q$
\EndFor
\EndProcedure
\end{algorithmic}
\end{algorithm}

\begin{table*}[t!]
    \parbox{.45\linewidth}{
    \centering
    \begin{tabular}{|c|c|}
    \hline
        \textbf{Type} & \textbf{Size}  
        \\
        \hline
        Input layer & $d \times d$\\
        \hline
        Convolutional layer & 512 filters, $3 \times 3$ size, $2\times 2$ stride \\
        \hline
        Convolutional layer & 256 filters, $3 \times 3$ size, $2\times 2$ stride \\
        \hline
        Convolutional layer & 256 filters, $3 \times 3$ size, $2\times 2$ stride \\
        \hline
        Fully-connected layer & 256 neurons\\
        \hline 
        Fully-connected layer & 128 neurons\\
        \hline
        Fully-connected layer & 64 neurons\\
        \hline
        Fully-connected layer & 32 neurons\\
        \hline
        Output layer & 4 outputs\\
        \hline
    \end{tabular}
    \caption{Summary of the deep neural network architecture.}
    \label{NN_architecture}
    }
    \hfill
    \parbox{.4\linewidth}{
    \centering
    \begin{tabular}{|c|c|}
    \hline
        \textbf{Parameter} & \textbf{Value}  \\
        \hline
        Input layer & $d \times d$\\
        \hline
        Batch size & 32\\
        \hline
        Maximum training steps & 1000\\
        \hline
        Memory buffer size & 3000000\\
        \hline
        Target update iterations & 500\\
        \hline
        Learning rate & 0.0001\\
        \hline
        $Beta_1$ & 0.9\\
        \hline
        $Beta_2$ & 0.999\\
        \hline
        Optimizer & Adam\\
        \hline
        Rewards & \begin{tabular}{@{}c@{}}+1000 if success \\0 if failure  \end{tabular}\\
      \hline
    \end{tabular}
    \caption{Summary of the common hyper-parameters for all agents.}
    \label{common_ hyper}
    }
\end{table*}

\begin{table*}[t!]
    \centering
    \begin{tabular}{|c|c|c|c|c|c|c|c|}
    \hline
        \textbf{Agent} & \textbf{$\epsilon$ initial} & \textbf{$\epsilon$ final} & \textbf{$\#$ iterations} & \textbf{$p$ initial} & \textbf{$p$ final} & \textbf{$\#$ iterations with  $p$ final} & $\gamma$\\
       
        \hline
        $d=3$ & 0.75 & 0.08 & 1000 & 0.01 & 0.10 & 300 & 0.9\\
        
        \hline
        $d=5$ & 0.75 & 0.08 & 2500 &0.01 & 0.10 & 1000 &0.95\\
        
        \hline
        $d=7$ & 0.75 & 0.08 & 6000 & 0.01 & 0.10 & 2000 & 0.95\\
        
        \hline
        $d=9$ & 0.5 & 0.1 & 8000 & 0.01 & 0.10 & 3500 & 0.95\\
        \hline
    \end{tabular}
    \caption{Summary of the different hyper-parameters for each agent.}
    \label{single_ hyper}
\end{table*}

\subsection*{Results}

We now present our results on RL decoders for uncorrelated noise on the toric code. Agents were trained on  grids of dimensions, 
$d = 3, 5, 7$ and $9$.  After training, we evaluated the decoders performance for different error probabilities $p$. We define the logical 
success probability as the proportion of syndromes decoded successfully. As for uncorrelated errors the MWPM algorithm is near-optimal, we also compare the RL agent's performance with MWPM.

\begin{figure}[ht!]
    \includegraphics[scale=0.6]{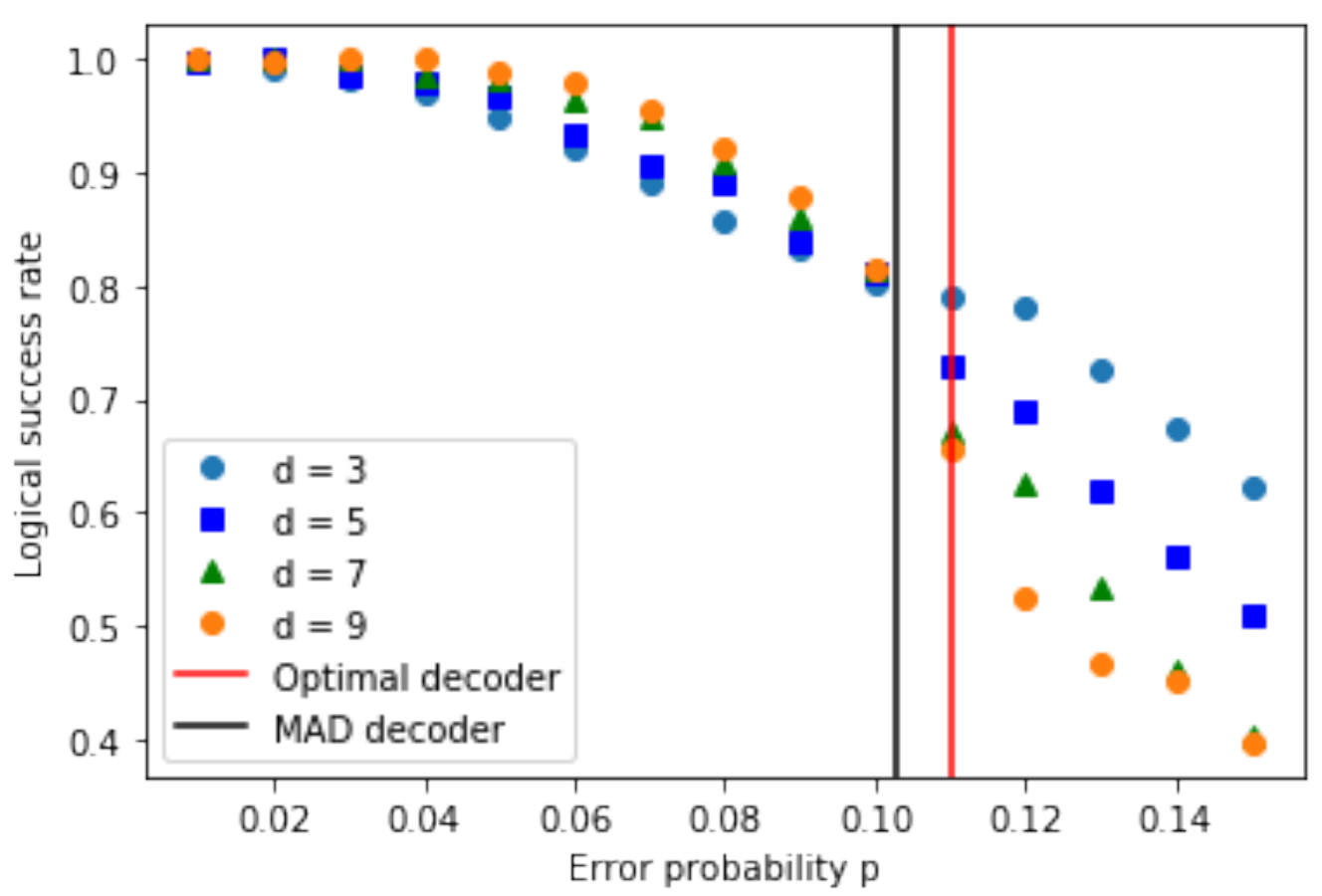}\; \includegraphics[scale=0.6]{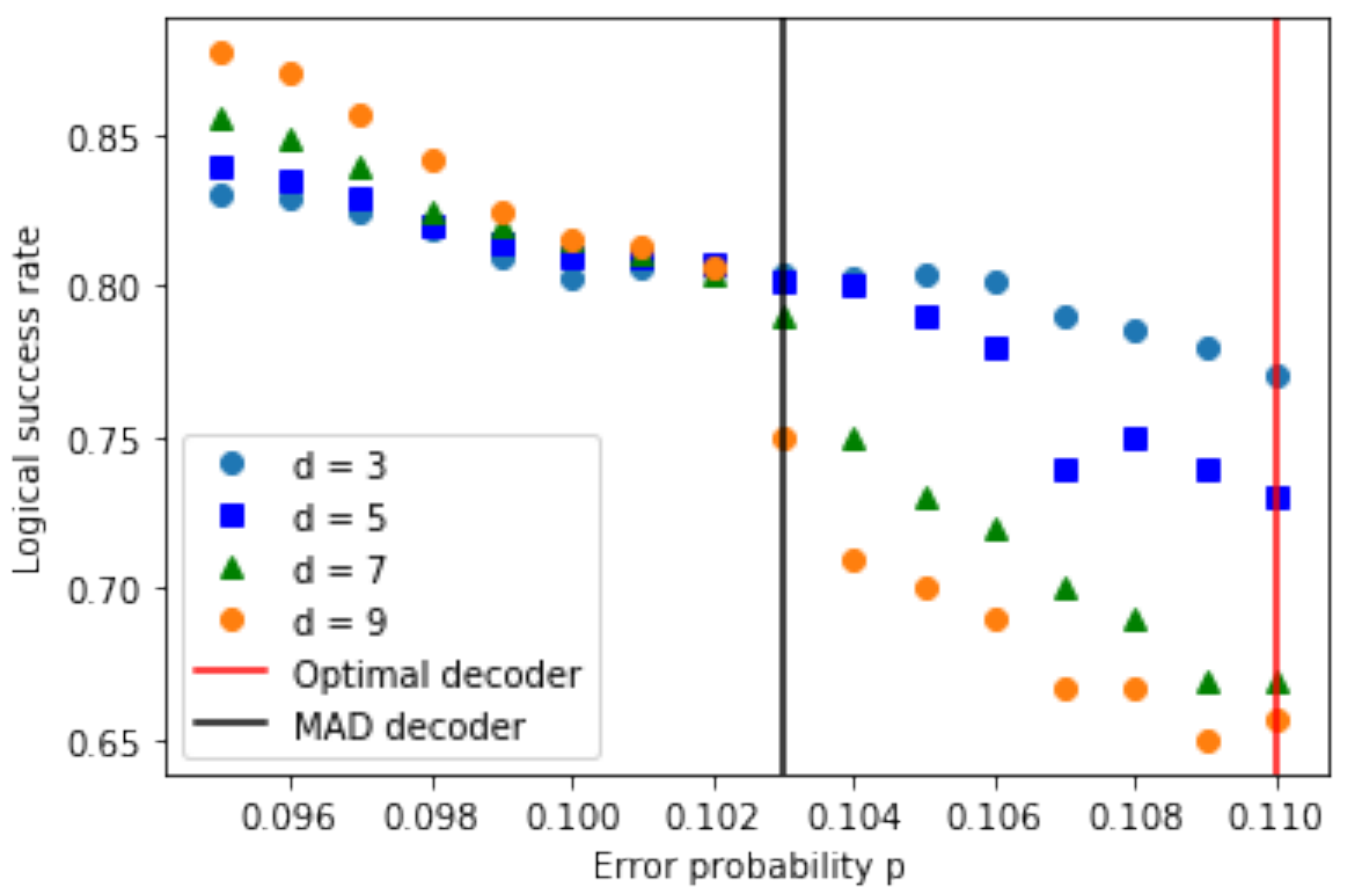}
    \caption{Logical success probability as a function of error probability $p$ for agents trained in different lattice dimensions $d=3,5,7,9$. The values of $p$ range from $0.01-0.15$ (top) and from $0.095-0.11$ (bottom). All agents were trained with error probability $p=0.10$. Thresholds for the optimal decoder and the MAD are shown as red and grey vertical bars, respectively.}
    \label{fig:bitflip}
\end{figure}

Fig.~\ref{fig:bitflip} shows the logical success probability as a function of the error probability $p$ for several lattice sizes 
For $p \leq 0.10$ the performance of the RL agent improves with the 
dimension $d$, of the lattice. When the error probability $p$ is low, the probability of successful decoding increases with the 
dimension of the lattice, whereas for $p$ high the opposite effect occurs.  The turning point between these two behaviors is 
called the \textit{code threshold}. The code threshold for these agents falls between $p=0.102$ and $p=0.103$, similar to those obtained 
in~\cite{Andreasson:19}. We also tested decoders which were trained with error probabilities 
higher than the code threshold $p=0.15,\,\mathrm{and}\, p=0.2$.  We noticed a slight increase in performance in the former case whilst 
in the latter decoders performed significantly worse.

\begin{figure}[h!]
\centering
    \includegraphics[scale=0.60]{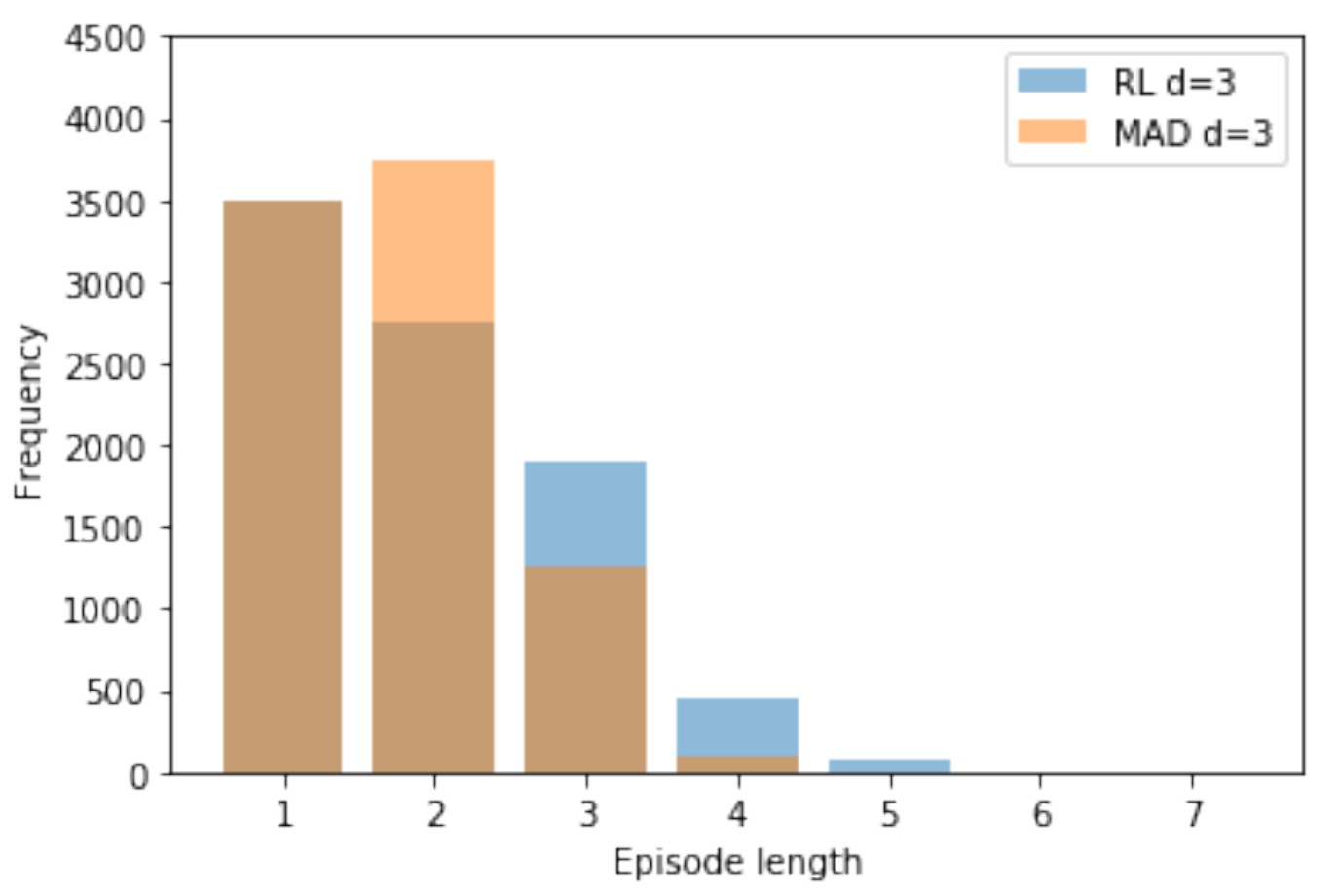}\;
    \includegraphics[scale=0.595]{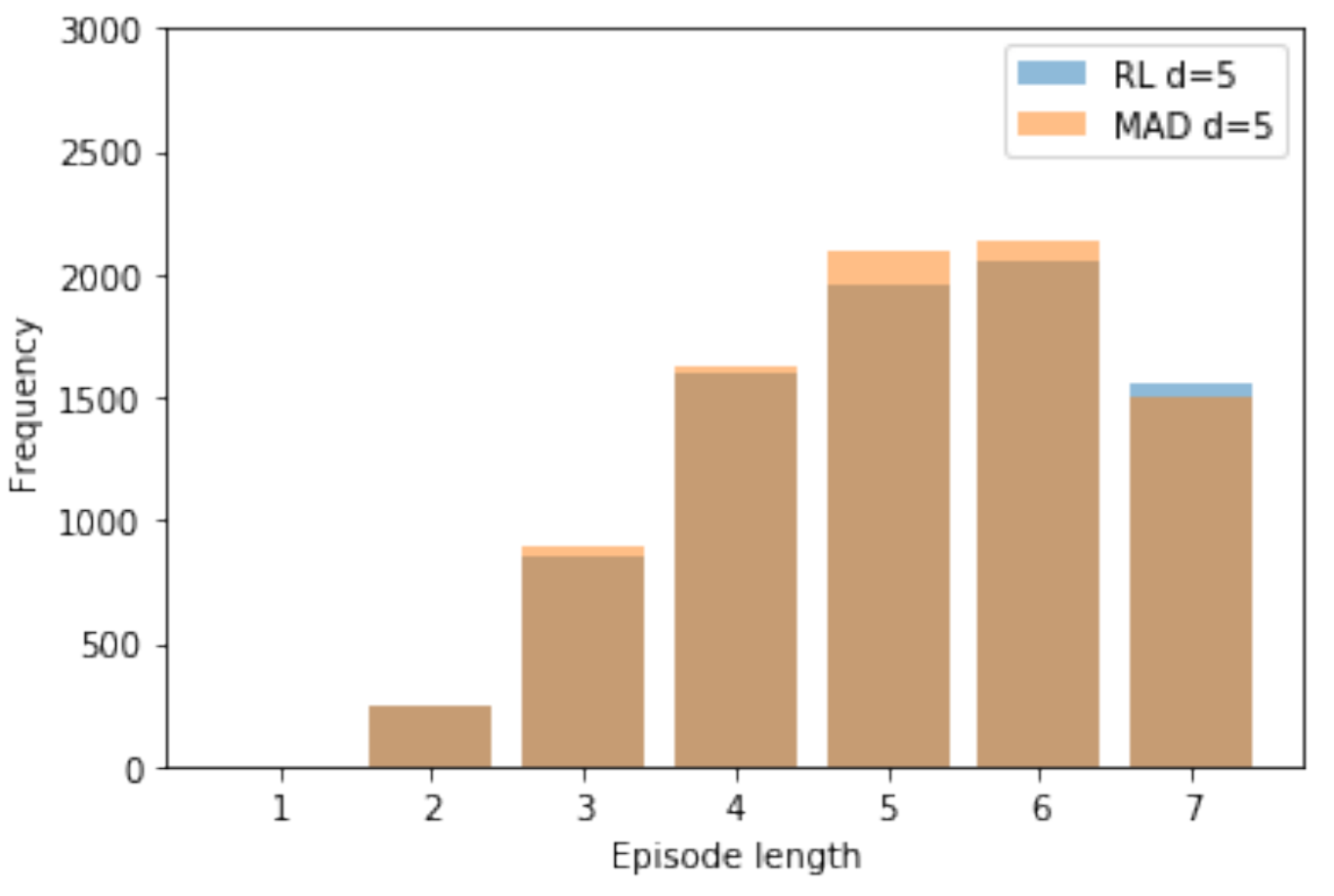}
     \caption {Episode length distribution for agents trained based on minimizing the number of actions and agents trained based on rewarding  solely on successful decoding for $d=3$ (left) and $d=5$ (right).}
     \label{fig:episode_lengths}
 \end{figure}

A more direct comparison between our decoder and the one of~\cite{Andreasson:19} is shown in Fig.~\ref{fig:episode_lengths}. We trained 
an agent according to the reward system of~\cite{Andreasson:19}, and compared the episode length distribution for both agents.
For $d=5$ both agents seem to have very similar episode length distributions meaning that an agent trained based on success/failure 
reward learnt that, in general, the best strategy is indeed to nullify the error syndrome as quickly as possible even though it was not explicitly 
told so. For $d=3$,  the  episode length distributions are more distinct.  By and large the agent adopts a decoding strategy requiring the least 
amount of corrections but occasionally slightly more steps are required.  

\begin{figure}[h!]
\centering
    \includegraphics[scale=0.6]{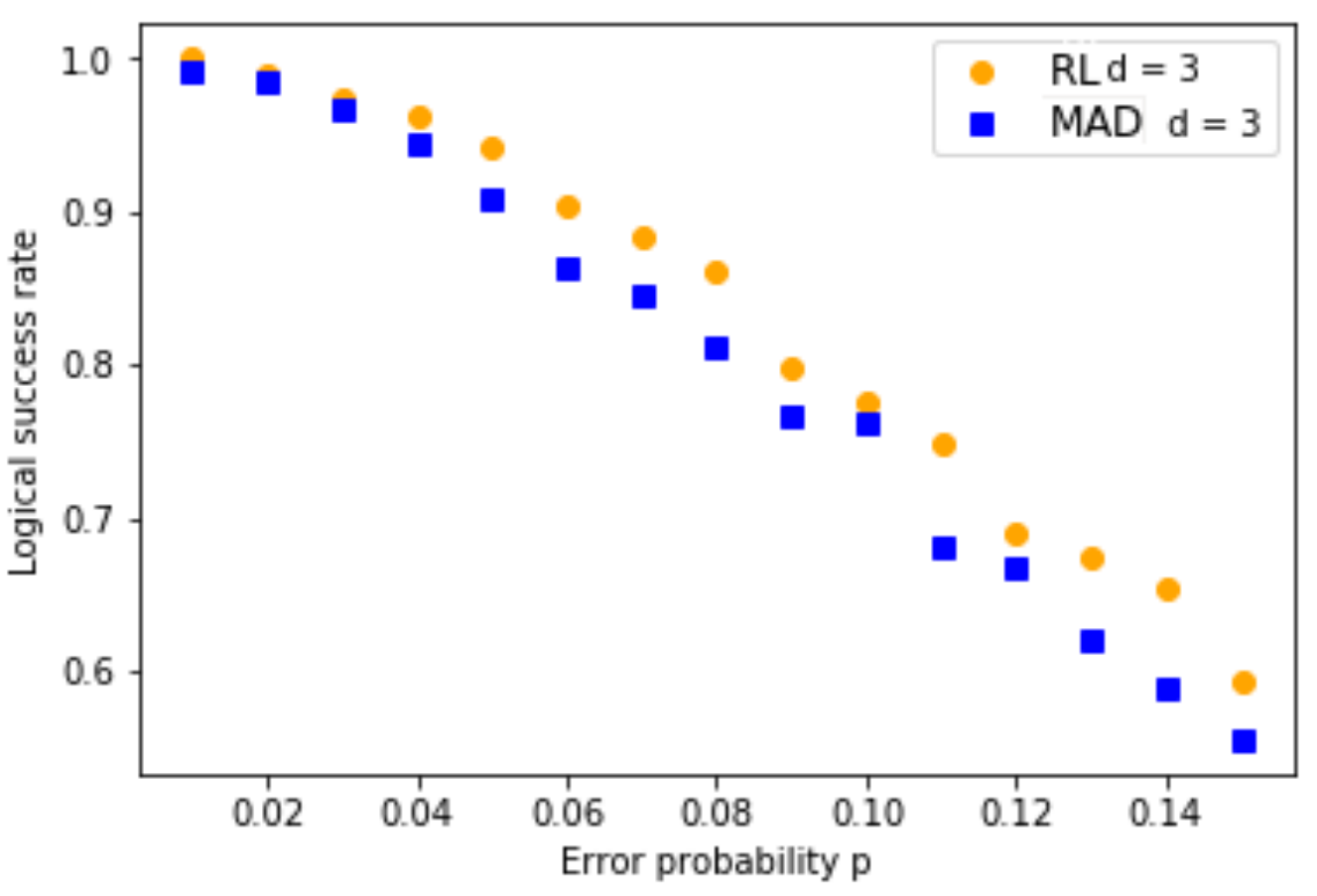}\;
    \includegraphics[scale=0.597]{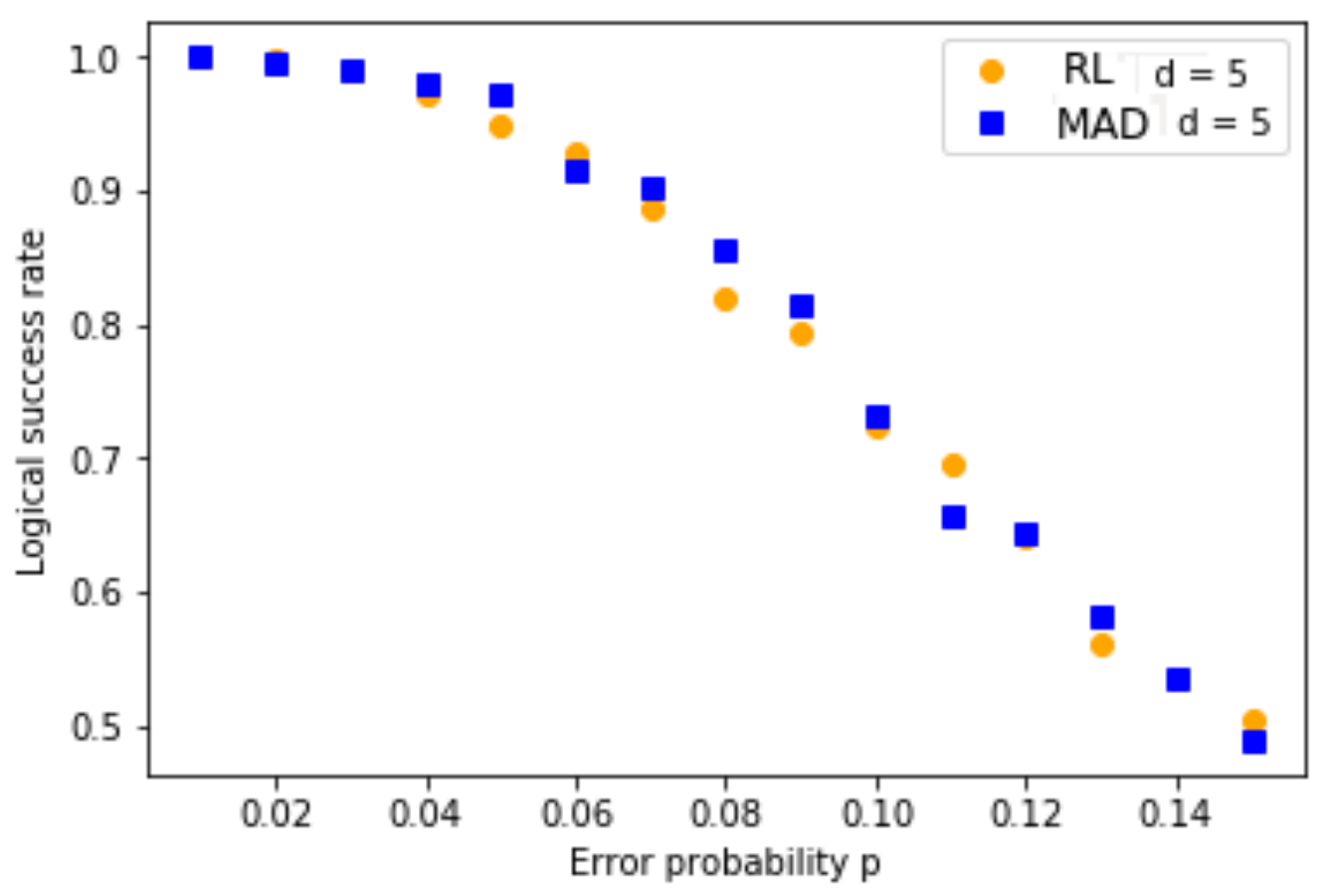}
     \caption {Logical success probability as a function of error probability $p$ for agents trained with success/failure rewards and rewards based on 
     the number of actions taken for $d=3$ (left) and $d=5$ (right).}
     \label{fig:efficiency_MWPM}
 \end{figure}

Our results indicate that it is not necessary to train agents based on minimizing their number of actions and that the natural choice of reward 
based on correct decoding is capable of learning the best policy, even yielding slight improvements in performance. A comparison of the 
efficiencies of the two decoders is shown in Fig. \ref{fig:efficiency_MWPM}. 
For $d=5$ both agents have very similar success probability. However, for $d=3$, the agent trained with success/failure 
rewards outperforms the one of~\cite{Andreasson:19} indicating that the different policy adopted by the success/failure agent is in fact 
beneficial for correct decoding of the code.

Finally, Table~\ref{time} shows the training times required for the various lattice sizes considered.  All training was performed using a single 
desktop computer.

\begin{table}[h!]
    \centering
    \begin{tabular}{|c|c|c|c|c|}
    \hline
        \textbf{Lattice dimension} & d=3 & d=5 & d=7 & d=9 \\
        \hline
        \textbf{Training time (hrs)} & 0.33 & 5 & 12 & 32\\
        \hline
    \end{tabular}
    \caption{Training time for each Reinforcement Learning model.}
    \label{time}
\end{table}

\section{Summary and Conclusions}
\label{sec:conclusion}
In this work we used deep Q-learning to train an agent using reinforcement learning in order to decode uncorrelated errors 
on a toric code.  We showed that agents can be trained based on whether or not correct decoding of the errors has been performed 
and showed that such decoders achieve near optimal performance.  Our agents are more versatile as compared to 
those trained based on minimising the number of actions taken, which leads to slightly improved performance on smaller lattices. 
Moreover, by comparing the episode length distributions between the two different types of agents we 
observe that indeed policies based on performing the minimum number of actions seem to form the most efficient decoders.

We believe that our model-free scheme of choosing rewards based solely on the success/failure of the 
decoding procedure is more versatile and can be used to design decoders for other TECC, such as surface codes, or the Kagome
lattice, as well as for more general noise models.  We expect that our approach is able to address error correction for correlated noise for which minimum action decoding methods are unsuitable. Work along this line is currently under consideration.

\section*{Acknowledgements}
The authors acknowledge support from Spanish MINECO reference FIS2016-80681-P (with the support of AEI/FEDER,EU); the 
Generalitat de Catalunya, project CIRIT 2017-SGR-1127 and the Baidu-UAB collaborative project 'Learning of Quantum Hidden 
Markov Models'.

\section*{References}
 \bibliographystyle{apsrev4-1}
\bibliography{RLearningToricCodes}
\end{document}